# $(NaCl)_x(KCl)_{y-x}(KBr)_{1-y}$ single crystals: study of the ac conductivity activation energy.

Vassiliki Katsika-Tsigourakou *

*Department of Solid State Physics, Faculty of Physics, University of Athens, Panepistimiopolis, 157 84 Zografos, Greece*

**Abstract**

Ac electrical measurements have been reported for alkali halide mixed crystals that were melt grown from NaCl, KCl and KBr starting materials. They showed a nonlinear variation of all the electrical parameters with the bulk composition. Here, we show that these mixed systems, depending on their major constituent, are classified into three categories in each of which, the activation energy for the ac conductivity increases linearly with $B\Omega$, where $B$ is the isothermal bulk modulus and $\Omega$ the mean volume per atom.

---

*vkatsik@phys.uoa.gr

## 1. Introduction

Since it was realized [1] that alkali halide mixed crystals have the potential for important applications in optical, optoelectronic and electronic devices, the interest on understanding their properties has been largely intensified, thus a large body of experimental studies have recently appeared [2-15]. This large amount of experimental data allowed the investigation of the correctness of early theoretical studies [16-20] the basic aim of which was to calculate the physical properties of a mixed system in terms of the corresponding properties of its end members. This recent investigation already gave some answers to some points summarized below.

First, Katsika-Tsigourakou and Vassilikou-Dova [21] showed that the compressibility $\kappa$ of the multiphased mixed crystals, grown by the melt method using the miscible NaBr and KCl, can be reliably estimated in terms of the elastic data of the end members alone. This made use of the compressibility of the defect formation volume calculated on the basis of a thermodynamical model [22-24], termed $cB\Omega$ model (see also below), that interconnects the Gibbs energy for the formation and migration of defects in solids with bulk properties. Subsequently, the same authors [25] showed that the large temperature variation of the static dielectric constant of the binary mixed crystal $(NaCl)_{0.587}(KBr)_{0.413}$ can quantitatively reproduced by using the interconnection between the ionic polarizability and the isothermal bulk modulus $B(=1/\kappa)$ proposed in Ref. [26]. In addition, in a later study [27], Katsika-Tsigourakou and Vassilikou-Dova showed that in four out of five mixed systems, formed when using NaBr and KCl as the starting materials, the activation energy for their conductivity process varies linearly with $B\Omega$, where $\Omega$ is the mean volume per atom. The same was found to hold [28] for the mixed crystals of NaBr and KBr that where also prepared from melt [29].

The aforementioned studies left open the question whether the $cB\Omega$ model [22-24, 30-32] conforms with experimental data when dealing with more complex mixed crystals, i.e., melt grown from three starting materials. Fortunately, such a case has appeared by Mahadevan and Jayakumari [33]. They presented electrical measurements on multiphased $(NaCl)_x(KCl)_{y-x}(KBr)_{1-y}$ single crystals that were melt grown from NaCl, KCl and KBr starting materials extending the investigation reported in $B$ Ref. [34]. They found a nonlinear variation of all electrical parameters they studied with the bulk composition. In view of this nonlinearity it is worthwhile to investigate in this short paper whether the linear interrelations between the physical properties that are predicted by the $cB\Omega$ model still hold in these multiphased $(NaCl)_x(KCl)_{y-x}(KBr)_{1-y}$ single crystals.

## 2. Investigation on the existence of a linear interrelation among the properties of $(NaCl)_x(KCl)_{y-x}(KBr)_{1-y}$ single crystals

The so called $cB\Omega$ model suggests that for the defect Gibbs energy $g^{act}$, for a given thermally activated process, is given by:

$$g^{act} = c^{act} B\Omega \tag{1}$$

where $c^{act}$ is a constant independent of temperature and pressure. The value of $c^{act}$, however, does depend on the host material as well as on the type of the defect process under investigation. By inserting Eq. (1) into the relation $s^{act} = -\left(dg^{act}/dT\right)_P$, where $s^{act}$ denotes the activation entropy, we can calculate the corresponding activation enthalpy $h^{act}$ through the relation $h^{act} = g^{act} + Ts^{act}$, which finally gives:

$$h^{act} = c^{act}\Omega\left[B - T\beta B - T\left(\frac{dB}{dT}\right)_P\right] \tag{2}$$

where $\beta$ is the thermal volume expansion coefficient. When the conductivity $\sigma$ varies with temperature in a way so that the plot $\ell n(\sigma T)$ vs $1/T$ is a straight line, the activation enthalpy $h^{act}$ coincides with the so called activation energy $E$ [35-37], thus:

$$E = h^{act} \tag{3}$$

The validity of Eqs (1) and (2) has been checked in various categories of solids, including the complex case of the defects that give rise to the emission of electric signals when ionic crystals are subjected to a time dependent stress. (This may provide the basis for the explanation [38] of the electric signals that precede earthquakes [39-43]). A review of a large variety of applications can be found in Ref. [44].

In the low temperature range, $Ts^{act}$ is significantly smaller than $h^{act}$, thus we can approximate $g^{act} \approx h^{act}$, and then Eqs (1) and (2) simplify to $h^{act} \approx c^{act}B\Omega$ which alternatively reads:

$$E = c^{act}B\Omega \tag{4}$$

In view of this expectation we plot in Fig. 1 the activation energies obtained from the ac conductivity measurements of Mahadevan and Jayakumari [33] versus $B\Omega$. The relevant values are given in Table 1, for all the 20 mixed systems that were melt grown from three (i.e., NaCl, KCl and KBr) starting materials studied in Ref. [33]. The values of $B\Omega$ are taken from the compressibilities $\kappa(=1/B)$ reported in Ref. [2] and the $\Omega$ values have been calculated as follows: We consider a mixed system $A_xB_{y-x}C_{1-y}$ where the three end members A, B and C will be equivalently labeled as the pure components (1), (2) and (3) and volumes $\upsilon_1$, $\upsilon_2$ and $\upsilon_3$ per "molecule" respectively [43]. We assume that the pure component (1) is the major component in the aforementioned mixed system $A_xB_{y-x}C_{1-y}$ and the volume $\upsilon_1$ per "molecule" is smaller than the volumes $\upsilon_2$ and $\upsilon_3$ of the pure components (2) and (3). Now, let $V_1 = N\upsilon_1$, $V_2 = N\upsilon_2$ and $V_3 = N\upsilon_3$ denote the corresponding molar volumes (where N stands for Avogadro's number). We define a "defect volume" $\upsilon_{2,1}^d$ as the increase in the volume $V_1$ if one molecule of type (1) is replaced by one molecule of type (2) and $\upsilon_{3,1}^d$ the increase in the volume $V_1$ if one molecule of type (1) is replaced by one molecule of type (3). Thus the addition of one molecule of type (2) to a crystal containing N molecules of type (1) will increase its volume by $\upsilon_{2,1}^d + \upsilon_1$, while an addition of one molecule of type (3) by $\upsilon_{3,1}^d + \upsilon_1$. If $\upsilon_{2,1}^d$ and $\upsilon_{3,1}^d$ are independent of composition, the volume $V_{N+n_2+n_3}$ of a crystal containing N molecules of type (1), $n_2$ molecules of type (2) and $n_3$ molecules of type (3) should be equal to: $V_{N+n_2+n_3} = N\upsilon_1 + n_2(\upsilon_{2,1}^d + \upsilon_1) + n_3(\upsilon_{3,1}^d + \upsilon_1)$. In the hard-spheres model, the defect volumes $\upsilon_{2,1}^d$ and $\upsilon_{3,1}^d$ can be approximately determined from: $\upsilon_{2,1}^d = \upsilon_2 - \upsilon_1$ and $\upsilon_{3,1}^d = \upsilon_3 - \upsilon_1$. Finally, the mean volume per atom $\Omega$ $(= V_{N+n_2+n_3}/N)$, is then calculated from: $\Omega = \upsilon_1 + (n_2/N)\upsilon_2 + (n_3/N)\upsilon_3$. $E_{ac}$

At first glance, an inspection of Fig. 1 does not reveal any systematic of $E_{ac}$ vs $B\Omega$. Disregarding, however the four points (i.e., numbers 4, 9, 11 and 17), that strongly deviate from the others, the remaining 16 points can be classified as follows: First, the four points 5, 6, 7 and 12 which correspond to the mixed systems that have NaCl as the major constituent. These are plotted in Fig. 2 and more or less follow a linear trend. A least squares fit to a straight line (which is drawn in Fig. 2) gives a slope 0.036 and an intercept $\approx$-0.23 eV. Repeating the same procedure for the four mixed systems that have KCl as a major constituent (Fig. 3), we find a straight line with slope 0.043 and an intercept -0.35 eV. Finally, the remaining eight systems that have KBr as a major constituent (Fig. 4), result in an almost linear behavior with slope 0.027 and an intercept -0.15 eV. Since the latter slope seems to be markedly smaller than those deduced in Figs 2 and 3, we further investigated that case and found the following: If we just disregard the point N$^o$ 20, all the other seven points in Fig. 4, i.e., N$^o$ 10, 13, 14, 15, 16, 18, and 19, lead to a slope 0.043 (and an intercept -0.48 eV) which is almost the same with the slopes obtained from Figs 2 and 3. This corresponds to the straight line drawn in Fig. 4. In other words, for the vast majority (i.e., seven out of eight) of the mixed systems that have KBr as a major constituent, we may say that they result in a straight line having a slope almost equal to that deduced from the mixed systems that have NaCl or KCl as a major constituent.

## 3. Discussion

The exact reason why in Fig. 1 some of the points deviate strongly from the others is not clear. A tentative reason could be guessed if we consider that the conductivity plot $\ell n(\sigma T)$ vs $1/T$ of alkali halides mainly consists of three almost linear segments as follows. Assuming for simplicity that the cation vacancies are appreciably more mobile than the anion vacancies, the segment corresponding to the higher temperature range (close to the melting point), termed intrinsic region, has a slope equal to the $\left(h^f/2 + h^m\right)/k_B$ where $h^f$ is the formation enthalpy per Schottky defect and $h^m$ the cation vacancy migration enthalpy. In the intermediate temperature range, termed extrinsic region, the slope is equal to $h^m/k_B$, while in the lowest temperature range, termed “association” region, the slope becomes larger, i.e. equal to $(h^m + h^{ass})/k_B$ where $h^{ass}$ is the dissociation enthalpy. This is so, because the crystal usually contains divalent cation impurities which, for the sake of charge neutrality, produce extrinsic cation vacancies that have negative effective charge. At low temperatures, a portion of these vacancies are attracted by the divalent impurities, thus forming electric dipoles, termed complexes. The quantity $h^{ass}$ is just the heat necessary to dissociate these complexes (thus giving free cation vacancies that contribute to conductivity).

Since the measurements of Mahadevan and Jayakumari [33] were carried out at temperature ranging from 308 to 423 K, the following might have happened. Some points in Fig. 1, may correspond to systems that are in the extrinsic region of their conductivity plot, while the others to the dissociation region. Hence, the former systems may have activation energies equal to $h^m$ while the others equal to $h^m + h^{ass}$, thus the $E_{ac}$ values plotted in Fig. 1 do not correspond to the same process and hence some of them naturally deviate significantly from the others. This is might also explain why one of the points (i.e., N$^o$ 20) in Fig. 4 does not fully conform with the straight line resulted from the other seven points, which –as mentioned- has almost the same slope with the straight lines drawn in Figs 2 and 3.

The following remark might be worthwhile to be added: The straight lines drawn in Figs 2, 3 and 4 do not pass through the origin of the axis, as demanded by Eq. (4). The exact reason is yet unknown, although it might be attributed in general to experimental uncertainties in the determination of $E_{ac}$.

## 4. Conclusion

Inspired from the so called $cB\Omega$ model, in this short paper, we studied the multiphased $(NaCl)_x(KCl)_{y-x}(KBr)_{1-y}$ single crystals. We find that, when considering mixed crystals that have the same major constituent, i.e., NaCl or KCl or KBr, the activation energy for the ac conductivity process seems to increase linearly versus $B\Omega$. It is challenging that the resulting three straight lines (comprising four, four and seven systems, respectively) have almost the same slope, i.e., around 0.04.

## FIGURE CAPTIONS

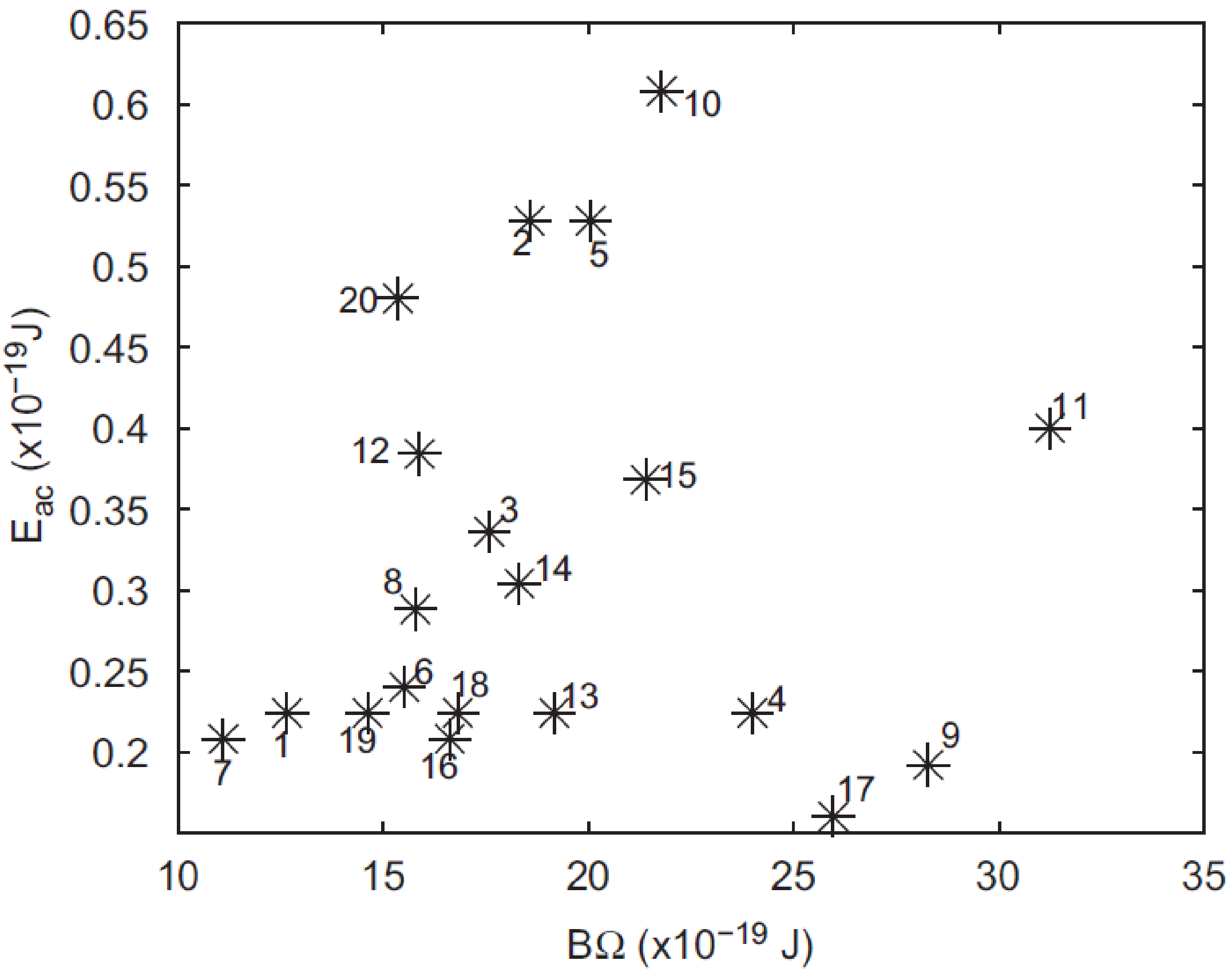

**Figure 1.** The activation energy for the ac conductivity, as measured by Mahadevan and Jayakumari [33], versus the quantity $B\Omega$ for all multiphased $(NaCl)_x(KCl)_{y-x}(KBr)_{1-y}$ single crystals. The numbers refer to those used in Table 1.

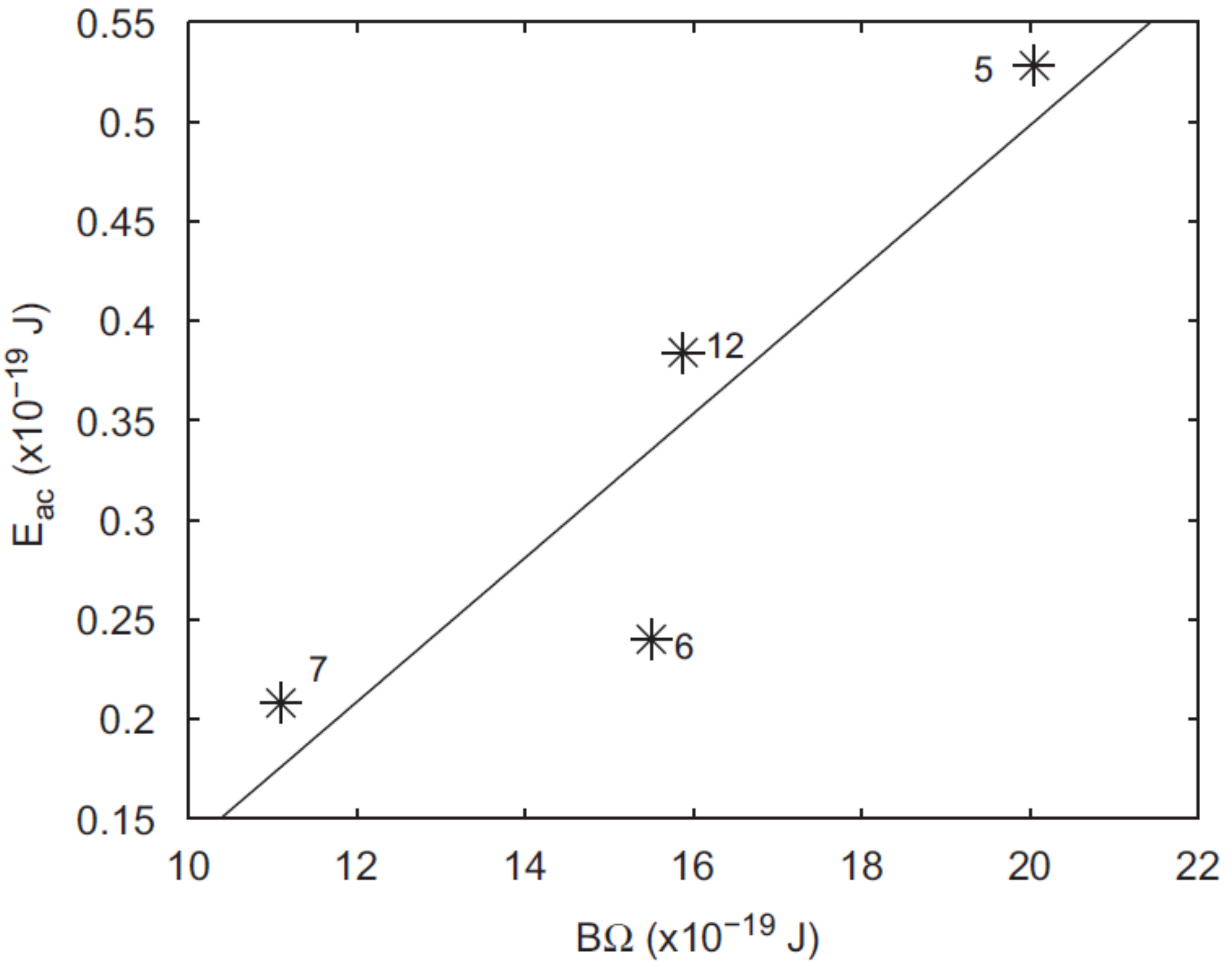

**Figure 2.** The activation energy for the ac conductivity, as measured by Mahadevan and Jayakumari [33], versus the quantity $B\Omega$ for those multiphased $(NaCl)_x(KCl)_{y-x}(KBr)_{1-y}$ single crystals for which NaCl is the major component. A least squares fit to a straight line results in the line drawn in the figure. The numbers refer to those used in Table 1.

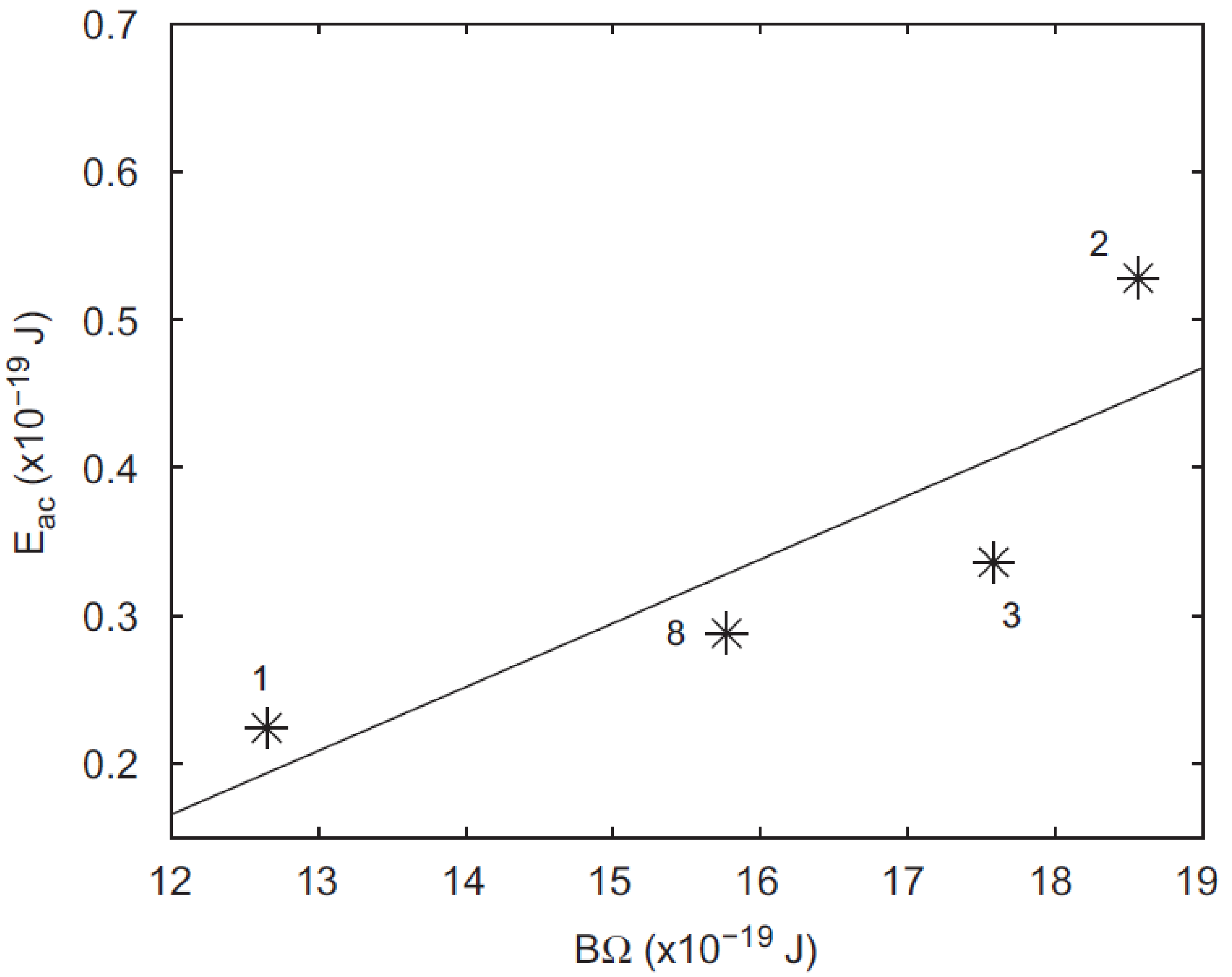

**Figure 3.** The activation energy for the ac conductivity, as measured by Mahadevan and Jayakumari [33], versus the quantity $B\Omega$ for those multiphased $(NaCl)_x(KCl)_{y-x}(KBr)_{1-y}$ single crystals for which KCl is the major component. A least squares fit to a straight line results in the line drawn in the figure. The numbers refer to those used in Table 1.

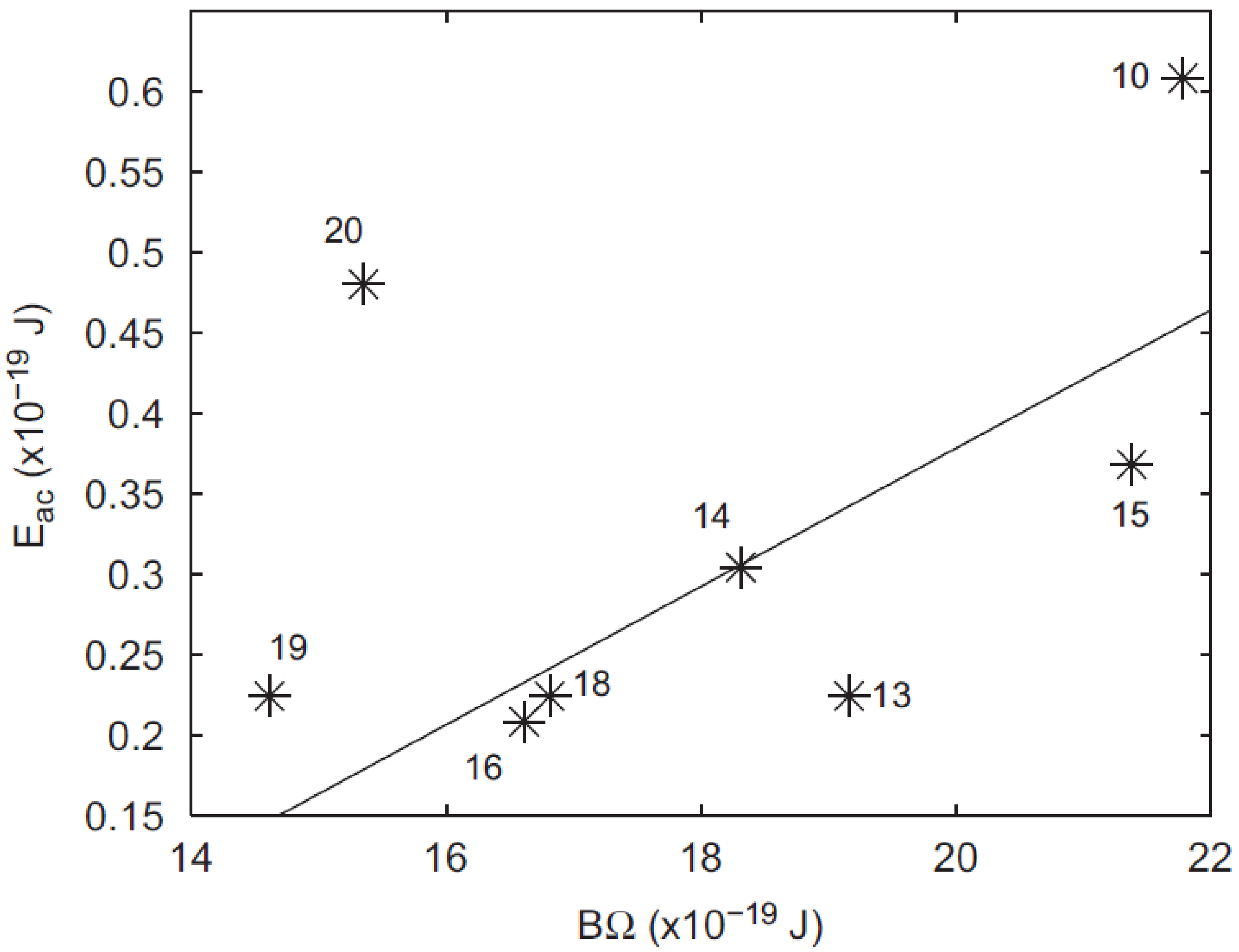

**Figure 4.** The activation energy for the ac conductivity, as measured by Mahadevan and Jayakumari [33], versus the quantity $B\Omega$ for those multiphased $(NaCl)_x(KCl)_{y-x}(KBr)_{1-y}$ single crystals for which KBr is the major component. The line drawn in the figure is found by a least squares fit to a straight line of the seven points 10, 13, 14, 15, 16, 18, and 19 (i.e., the point N$^o$ 20 has been disregarded, see the text). The numbers refer to those used in Table 1.

**Table 1.** The values of the activation energy for the ac conductivity $E_{ac}$ (as reported in Ref. [33]), the compressibility $\kappa$ (as reported in Ref. [2]) and $B\Omega$ for the $(NaCl)_x(KCl)_{y-x}(KBr)_{1-y}$ single crystals. The values of $B$ are taken from the compressibilities $\kappa(=1/B)$ reported in Ref. [2] and the $\Omega$ values are calculated with the procedure explained in the text.

| System[a] | $E_{ac}$ $10^{-19}$ J | $\kappa$ $(10^{-11}\ m^2/N)$ | $B\Omega$ $10^{-19}$ J |
|---|---|---|---|
| **1**: $(NaCl)_{0.078}(KCl)_{0.724}(KBr)_{0.198}$ | 0.224 | 6.859 | 12.65 |
| **2**: $(NaCl)_{0.159}(KCl)_{0.641}(KBr)_{0.200}$ | 0.528 | 5.163 | 18.56 |
| **3**: $(NaCl)_{0.282}(KCl)_{0.524}(KBr)_{0.194}$ | 0.336 | 6.429 | 17.58 |
| **4**: $(NaCl)_{0.389}(KCl)_{0.418}(KBr)_{0.193}$ | 0.224 | 5.720 | 23.99 |
| **5**: $(NaCl)_{0.479}(KCl)_{0.319}(KBr)_{0.202}$ | 0.528 | 5.816 | 20.05 |
| **6**: $(NaCl)_{0.595}(KCl)_{0.218}(KBr)_{0.187}$ | 0.240 | 5.822 | 15.50 |
| **7**: $(NaCl)_{0.704}(KCl)_{0.091}(KBr)_{0.205}$ | 0.208 | 6.651 | 11.10 |
| **8**: $(NaCl)_{0.063}(KCl)_{0.541}(KBr)_{0.396}$ | 0.288 | 7.618 | 15.77 |
| **9**: $(NaCl)_{0.159}(KCl)_{0.453}(KBr)_{0.388}$ | 0.192 | 4.938 | 28.27 |
| **10**: $(NaCl)_{0.292}(KCl)_{0.029}(KBr)_{0.419}$ | 0.608 | 4.932 | 21.78 |
| **11**: $(NaCl)_{0.361}(KCl)_{0.212}(KBr)_{0.427}$ | 0.400 | 4.504 | 31.24 |
| **12**: $(NaCl)_{0.505}(KCl)_{0.039}(KBr)_{0.457}$ | 0.384 | 7.222 | 15.88 |
| **13**: $(NaCl)_{0.133}(KCl)_{0.363}(KBr)_{0.504}$ | 0.224 | 6.709 | 19.16 |
| **14**: $(NaCl)_{0.230}(KCl)_{0.274}(KBr)_{0.493}$ | 0.304 | 6.957 | 18.31 |
| **15**: $(NaCl)_{0.261}(KCl)_{0.231}(KBr)_{0.508}$ | 0.368 | 5.763 | 21.38 |
| **16**: $(NaCl)_{0.389}(KCl)_{0.075}(KBr)_{0.536}$ | 0.208 | 6.812 | 16.61 |
| **17**: $(NaCl)_{0.110}(KCl)_{0.293}(KBr)_{0.597}$ | 0.160 | 4.265 | 25.96 |
| **18**: $(NaCl)_{0.240}(KCl)_{0.159}(KBr)_{0.602}$ | 0.224 | 6.316 | 16.81 |
| **19**: $(NaCl)_{0.272}(KCl)_{0.103}(KBr)_{0.625}$ | 0.224 | 6.958 | 14.61 |
| **20**: $(NaCl)_{0.104}(KCl)_{0.079}(KBr)_{0.817}$ | 0.480 | 5.446 | 15.35 |

[a]estimated bulk composition in the crystal [33]